
\magnification=\magstep1
\baselineskip=16pt
\def\lanbox{\hbox{$\, \vrule height 0.25cm width 0.25cm depth 0.01cm \,$}}

\input amssym.def     

\font\eightit=cmti8
\font\eightpoint=cmr8


\hbox to \hsize{\hfil\eightit 15/September/1999}
\vglue1truein
\centerline{\bf A SIMPLE PROOF OF A THEOREM OF}
\centerline{\bf LAPTEV AND WEIDL}

\bigskip
\bigskip
\bigskip 

{\baselineskip=2.5ex
\vfootnote{}{\eightpoint
\noindent\copyright 1999 by the authors.
Reproduction of this article, in its entirety, by any means is permitted
for non-commercial purposes.}}

{\baselineskip = 12pt
{\qquad \qquad  Rafael Benguria$^*$   \qquad \qquad \qquad \qquad \qquad
Michael Loss $^{**}$ }

{ \  \ \quad \qquad Facultad de Fisica\  \qquad \qquad \qquad  \quad 
School of Mathematics} 

{\ \  \qquad P. U. Cat\'olica de Chile \qquad  \qquad 
Georgia Institute of Technology} 

{\  \ \quad  Casilla 306, Santiago 22, Chile \  \ \qquad  \quad  Atlanta, GA
30332, USA }

\vfootnote{}{$^*$ Work supported by Fondecyt (Chile) project 199--0427 and by a
John Simon Guggenheim Memorial Foundation fellowship, $^{**}$ Work supported by
N.S.F. grant DMS-95-00840} }
\vskip 1 true in
{\bf Abstract:}

A new and elementary proof of a recent result of Laptev and Weidl [LW] is given.
It is a sharp Lieb--Thirring inequality for one dimensional Schr\"odinger
operators with matrix valued potentials.

\vfill\eject
{\bf I. Introduction}

In this note we give a new and, we believe, simpler   
proof of a recent 
result of Laptev and Weidl. It is concerned with Lieb--Thirring
inequalities for matrix
valued Schr\"odinger operators of the type
$$
H =- {{\rm d}^2 \over {\rm d}x^2} \otimes I +V(x) \eqno(1)
$$ 
acting on $L^2(\Bbb{R} ;\Bbb{C}^N)$. The potential $V(x)$ is a
negative definite hermitean 
$N\times N$ matrix. We assume that its matrix elements are smooth functions
of compact support, say in the interval $[-a,a]$. The operator $H$ has finitely
many negative eigenvalues, which, counting multiplicities, we denote  by 
$-\lambda_j$
$j=1,
\dots, L$.

 The following theorem was proved in [LW].
\medskip
{\bf Theorem 1}

{\it With the above assumptions on $V$ the following inequality holds
$$
\sum_{j=1}^L\lambda_j^{3/2} \leq {3 \over 16} \int_{\Bbb{R}} {\rm Tr}(V(x)^2) {\rm d}x
\ .\eqno(2)
$$
}
\medskip
{}From Weyl's law on the distribution of eigenvalues it is seen that this
inequality is best possible. For the case where the potential is a scalar
function this result was already proved in [LT] where it was realized that
(2) follows from trace identities.

The matrix case, however, is important, since inequality  (2) of 
Laptev and Weidl
is the starting point for deriving sharp Lieb--Thirring inequalities in higher
dimensions. In particular, the argument of [AL] applies also in this case
and yields sharp Lieb-Thirring inequalities for the sum of powers of eigenvalues
where the power is larger than $3/2$.
For the details we refer the reader to the original paper [LW] where a
collection of beautiful results is presented. 
Their proof of Theorem 1 which corresponds to formula (2.1) in their paper is 
patterned after the proof of [BF] (see also [FZ]) and 
is fairly involved.
Inequality (2) is derived from a trace identity, which in turn
is a special case of a whole family of identities  that express
conservation laws of the Korteweg--de Vries equation. 
The derivation of these trace identities uses nontrivial results
about scattering theory on
the line and Laptev and Weidl prove these afresh for the matrix case. 
Since (2) is the central result in [LW] and of independent interest, 
it is of value to have a different,  more elementary
and more direct proof.  It relies on the
`commutation method' and some
elementary facts from the calculus of variations. 

The `commutation method' has a
fairly long history, some versions of it were already known to Darboux [DG] and
Jacobi[J]. Its modern appearance seems to be due to Crum [C]. For a rigorous 
discussion of these issues we refer to the papers of [G] and [DP]. In the latter
more examples of the usefulness of this method are presented.
Another work, closer to the spirit of ours, is the one of Schmincke [S] who 
uses the commutation method to prove that
$$
\sum \lambda_j^{1/2} \geq -{1 \over 4} \int V(x) {\rm d} x
$$
for scalar potentials. This result was extended in [LW] to the matrix case
which can also be obtained using the methods of the present work.
This inequality should be contrasted with 
$$
\sum \lambda_j^{1/2} \leq -{1 \over 2} \int V(x) {\rm d} x
$$
obtained in [HLT] for the scalar case and in [HLW] for the matrix case.
Both inequalities are sharp in the sense that the constants cannot be improved.

To illustrate the ideas we give a short proof of Theorem 1 for the 
case where $V$ is a scalar potential, thereby recovering the
result in [LT]. This sets the stage
for the proof of the matrix  case in the following section.
While it is certainly possible to prove Theorem 1 under fairly general conditions 
on the potential, we refrain from doing so. It would clutter the simple argument
with  technical details. 

\vskip .5 true in
{\bf II. The scalar case}

Let $-\lambda_1$ be the lowest eigenvalue of the Schr\"odinger operator
(1) of Section I with a {\it scalar} potential. It is well known that this eigenvalue
is not degenerate and the
corresponding eigenfunction $\phi_1$ can be chosen to be strictly positive.
Moreover, outside the range of the potential we have
$$
\phi_1(x) = \cases{ 
{\rm  const.}e^{-\sqrt{\lambda_1}x},& if $x>a$,\cr
{\rm  const.}e^{\sqrt{\lambda_1}x},& if $x<-a$.\cr 
}\eqno(1)
$$
Thus the function
$$
F(x) = {\phi_1^{\prime}(x) \over \phi_1(x)} \ , \eqno(2)
$$
is defined and satisfies the Riccati equation
$$
F^{\prime} + F^2 = V+\lambda_1 \ ,\eqno(3)
$$
together with the conditions
$$
F(x)=\cases{ -\sqrt{\lambda_1},& if $x>a$, \cr \sqrt{\lambda_1},& 
if $x<-a$. \cr} \eqno(4)
$$
A simple computation shows that the Hamiltonian $H$ can be written as
$$
H=D^{*} D - \lambda_1 \ ,\eqno(5)
$$
where
$$
D={{\rm d} \over {\rm d}x} -F\ ,\eqno(6)
$$
and 
$$
D^{*}=-{{\rm d} \over {\rm d}x} -F \ .\eqno(7)
$$
It is a general fact [DP][G] that the operators $D^{*}D$ and $DD^{*}$ on
$L^2(\Bbb{R})$ have the
same spectrum with the possible exception of the zero eigenvalue. Note
that $D^{*}D$ has a zero eigenvalue which corresponds to the ground state
of $H$. The operator $DD^{*}$ does not have a zero eigenvalue. This follows
from the fact that the corresponding eigenfunction $\psi$ satisfies
$$
\psi^{\prime} = -F \psi \ ,\eqno(8)
$$
and hence $\psi(x) = {\rm const.}/ \phi_1(x)$ which grows exponentially 
and is not normalizable.
Thus the new Schr\"odinger operator
$$
\tilde H = DD^{*} -\lambda_1 = -{{\rm d}^2 \over {\rm d} x^2} -F^{\prime} +F^2
-\lambda_1
= -{{\rm d}^2 \over {\rm d} x^2} +V -2F^{\prime}\ .\eqno(9)
$$
has, except for the eigenvalue $-\lambda_1$, precisely the same eigenvalues
as $H$. Also note that the potential $V-2F^{\prime}$ is smooth and has 
support in the same interval as the potential $V$.

Next, we compute using the Riccati equation (3)
$$
\int(V-2F^{\prime})^2 {\rm d} x =  \int V^2 {\rm d} x +4\int(\lambda_1
-F^2)F^{\prime}{\rm d} x \ .
$$
The last term can be computed explicitly using (4) and we obtain
$$
\int(V-2F^{\prime})^2 {\rm d} x= \int V^2 {\rm d} x - {16 \over 3}
\lambda_1^{3/2} \ . \eqno(10)
$$
Thus,
$$
\sum_{k=1}^L \lambda_k^{3/2} - {3 \over 16} \int V^2 {\rm d} x
= \sum_{k=2}^L \lambda_k^{3/2} - {3 \over 16} \int(V-2 F^{\prime})^2 {\rm d} x
\ ,\eqno(11)
$$
 and the Schr\"odinger
operator with the potential $V-2F'$ has precisely the eigenvalues
$-\lambda_2, \dots, -\lambda_L$. Continuing this process we remove one
eigenvalue after another. After the last one is removed a
manifestly negative quantity is left over, and this proves Theorem 1 in
the scalar case. \hfill\lanbox

\vskip .5 true in

{\bf III. The matrix case}
\bigskip
The proof of Theorem 1 is patterned after the scalar case. In addition
to the usual eigenvalue equation for $H$ in (1) of Section I
$$
-\phi ^{\prime \prime}(x) +V(x) \phi(x) = -\lambda \phi(x) \eqno(1)
$$
we consider the following matrix version for an $N \times N$ matrix $M(x)$,
$$
-M^{\prime \prime}(x) + V(x) M(x) = -\lambda M(x) \eqno(2)
$$
The following Lemma is central.
\bigskip
{\bf Lemma 2}
\medskip
{\it Assume that $-\lambda$ is the ground state energy of $H$ and let
$\phi$ be any solution of the differential equation (1) with
$$
\phi(x) = e^{\sqrt{\lambda}x} u\ \ {\rm for}\ \ x<-a
$$
where $0 \not= u \in \Bbb{C}^n$ is constant. In particular, we do not require that
$\phi$ is normalizable.
Then $\phi(x)$ never vanishes. Moreover, the ground state energy is at most
$N$--fold degenerate.}
\bigskip
{\it Proof:} Suppose there exists a point $x_0$ with $\phi(x_0)=0$.
Consider the continuous function
$$
\tilde{\phi}(x) =\cases{ \phi(x), &if $x < x_0$\cr 0, &if $x \geq x_0$. \cr}
$$
Clearly, this function does not vanish identically and is square integrable.
A simple integration by parts calculation shows that
$$
(\tilde{\phi}, H \tilde{\phi}) = -\lambda(\tilde{\phi},\tilde{\phi})\ ,
$$
and thus $ \tilde{\phi}$ is a ground state and must be  a solution of
the Schr\"odinger equation (1) which is an ordinary differential equation of
second order. Here $(\ ,\ )$ denotes the inner product on 
$L^2(\Bbb{R},\Bbb{C}^N)$. Since
$\tilde{\phi}$ vanishes to the right of $x_0$ the solution must vanish 
everywhere,
which is a contradiction. The  last
statement of the lemma is an immediate consequence of this. \hfill\lanbox
\bigskip
{\it Remark:} The above Lemma clearly generalizes to potentials that
do not have compact support but decay, e.g., exponentially, at infinity.
\bigskip
Consider any matrix solution $M(x)$ of the differential equation (2) subject 
to the condition
$$
M(x) = e^{\sqrt{\lambda}x} A \ \ {\rm for}\ \ x <-a\ ,\eqno(3)
$$
where $A$ is a {\it nonsingular} matrix.
By the previous Lemma 1, any solution of (1) that decays exponentially must be a
linear combination of the column vectors of $M(x)$. In particular, the ground
states themselves must be linear combinations of the column vectors of $M(x)$.
Also by Lemma 2 we know that the matrix $M(x)$ must be invertible
for every $x \in \Bbb{R}$. Hence it makes sense
to define
$$
F(x) = M^{-1}(x) M^{\prime}(x)  \ .\eqno(4) 
$$

The following Lemma 3 states all we need to know about $F(x)$. The number
$K$ below denotes the degeneracy of the ground state energy. We have that
$K\leq N$ by Lemma 2.
\bigskip
{\bf Lemma 3}
\medskip
{\it The matrix $F(x)$ is hermitean for every  $x \in \Bbb{R}$, independent of
the choice of $A$ and satisfies the matrix Riccati equation
$$
F^{\prime} + F^2 - V= \lambda I \ .\eqno(5)
$$
Moreover, for $x < -a$
$$
F(x) = \sqrt{\lambda}I \ ,\eqno(6)
$$
and for $x>a$, the eigenvectors of $F(x)$ are independent of $x$ and 
and its eigenvalues decay exponentially fast to the $K$ fold eigenvalue
$-\sqrt{\lambda}$ and the $N-K$ fold eigenvalue $\sqrt{\lambda}$ respectively}.
\bigskip
{\it Proof:} Consider any two matrix solutions of (2), $M_1(x)$ and $M_2(x)$.
{}From the Wronskian identity  ( with $*$ denoting adjoint)
$$
{{\rm d} \over {\rm d} x} \left(M_1(x)^{*}M_2^{ \prime}(x) -
M_1^{*\prime}(x)M_2(x) \right) = 0 \ ,\eqno(7)
$$
one obtains
$$
\left(M_1(x)^{*}M_2^{\prime}(x) -
M_1^{* \prime}(x)M_2(x) \right) = {\rm const.} \ .\eqno(8)
$$
First we set $M_1 =M_2=M$ and assuming the initial condition (3) we get that
$$
M(x)^{*}M^{\prime}(x) =
M^{*\prime}(x)M(x)  \ ,\eqno(9)
$$
which yields the hermicity of $F$. If we set $F_1
=M_1^{\prime}M_1^{-1}$ and $F_2 =M_2^{\prime}M_2^{-1}$ where $M_1$ and $M_2$
satisfy (2) and (3) for possibly two different, nonsingular matrices $A_1$ and
$A_2$ we get from (8) that
$F_1 \equiv F_2$. An elementary computation yields (5) and (6).

Fix $x_0 > a$. For $x>x_0$ the potential vanishes and the matrix $M(x)$ is given by
$$
M(x)= \cosh(\sqrt{\lambda}(x-x_0)) M(x_0) + {1 \over \sqrt{\lambda}}
\sinh(\sqrt{\lambda}(x-x_0))M^{\prime}(x_0) \ , \eqno(10)
$$
and hence $F(x)$ is given by
$$
F(x) = \sqrt{\lambda}\left(\sqrt{\lambda}\tanh(\sqrt{\lambda}(x-x_0))I + 
F(x_0)\right)\left( \sqrt{\lambda}I + \tanh(\sqrt{\lambda}(x-x_0))
F(x_0) \right)^{-1} \ . \eqno(11)
$$

{}From this it follows that the eigenvectors of $F(x)$ do not depend on $x$
and since $F(x)$ exists for all $x$ we must have $ -\sqrt{\lambda}I \leq
F(x_0) \leq \sqrt{\lambda}I$.
It follows from (10) that the bound states are precisely those solutions
$\phi(x)$ of the differential equation (1) that decay exponentially in both 
directions and that are of the 
form $\phi(x) = M(x)M^{-1}(x_0)u$ where $u$ is
an eigenvector of $F(x_0)$ with eigenvalue  $-\sqrt{\lambda}$.
Thus $F(x_0)$ has the $K$ fold eigenvalue $-\sqrt{\lambda}$ and all
the other eigenvalues $\nu_j$ satisfy the inequality $-\sqrt{\lambda}
< \nu_j \leq \sqrt{\lambda}$. 
{}From (11) we see $-\sqrt{\lambda}$ is a $K$ fold degenerate eigenvalue
of $F(x)$ for all $x \geq x_0$ and that all the other eigenvalues converge 
exponentially fast to $\sqrt{\lambda}$.\hfill\lanbox

\bigskip
{\it Proof of Theorem 1 in the matrix case:}
{}From the Riccati equation (5) we get that 
$$
H+\lambda_1 I = D^{*} D \ , \eqno(13)
$$
where
$$
D^{*}=\left(-{{\rm d} \over {\rm d}x}\otimes I-F\right)\ , \eqno(14)
$$
and
$$
D= \left({{\rm d} \over {\rm d}x}\otimes I -F \right)\ .\eqno(15)
$$
Clearly
$$
D\phi =0 \eqno(16)
$$
for any ground state $\phi$.
Moreover,
$$
D^{*} \psi =0 \eqno(17)
$$
has no nontrivial normalizable solution on $\Bbb{R}$ since
$F=\sqrt{\lambda} I $ for $x <-a$.

Thus the operator 
$$
H^{\prime} :=DD^{*}-\lambda_1 I \eqno(18)
$$
has precisely the eigenvalues $\lambda_{K+1}, \dots, \lambda_L$.
A calculation shows that
$$
H^{\prime}=  -{{\rm d}^2 \over {\rm d}x^2}\otimes I +V(x) -2F^{\prime}(x)
\eqno(19)
$$
where the potential
$$
V(x) -2F^{\prime}(x) \eqno(20)
$$
is  smooth and decays exponentially fast at infinity by Lemma 3.
One easily computes using (5) that
$$
\int_{\Bbb{R}} {\rm Tr}\left((V -2F^{\prime})^2 \right)  {\rm d}x
$$
$$
= \int_{\Bbb{R}} {\rm Tr}\left(V^2\right)  {\rm d}x
-4\int_{\Bbb{R}}
{\rm Tr}\left(\left(F^2-\lambda_1 I\right)F^{\prime}\right)  {\rm d}x \ ,
\eqno(21)
$$
which can be integrated to yield
$$
-{4 \over 3} {\rm Tr}\left(F^3(x) \right)|^{+\infty}_{-\infty}
+4\lambda_1{\rm Tr}\left(F(x)\right) |^{+\infty}_{-\infty} \ . \eqno(22) 
$$
By Lemma 2 this  equals
$$
 -{16 \over 3} K \lambda_1^{3/2} \ .\eqno(23)
$$
Again,  we have shown that
$$
\sum_{j=1}^L\lambda_j^{3/2} - {3 \over 16} \int_{\Bbb{R}} {\rm Tr}(V^2) {\rm d}x
= \sum_{j=K+1}^L\lambda_j^{3/2} - {3 \over 16} \int_{\Bbb{R}}
{\rm Tr}\left((V-2F^{\prime})^2\right) {\rm d}x\ ,\eqno(24)
$$
and the Schr\"odinger operator with the potential $V-2F^{\prime}$ has precisely
the eigenvalues $\lambda_{K+1}, \dots, \lambda_L$.

The potential $V-2F^{\prime}$ decays exponentially but, unfortunately, does
not have compact support and hence the second step in the scalar case, i.e., 
the removal of the next eigenvalue, cannot be
taken directly. However, the following approximation argument can be used
to circumvent this difficulty.
Cutting off the potential $V-2F^{\prime}$ sufficiently far
out we are left with a new potential $V-2F^{\prime}_c$ 
which has compact support
and whose eigenvalues are numbers $-\mu_{K+1}, \cdots
-\mu_L$ which can be made to be as close to the old ones  $-\lambda_{K+1}, \cdots
-\lambda_L$ as we please. The cutoff might cause some new eigenvalues to appear,
but all of those can be made to be as close to the continuum, i.e., as close 
to $0$ as we please.
Removing the ground state eigenvalue of this 
new potential $V-2F^{\prime}_c$ yields
$$
\sum_{j=K+1}^L \mu_j^{3/2} - {3 \over 16}\int
{\rm Tr}\left((V-2F^{\prime}_c)^2\right) {\rm d}x =
$$
$$
\sum_{j=K+P+1}^L \mu_j^{3/2} - {3 \over 16}\int
{\rm Tr}\left((V-2F^{\prime}_c -2G^{\prime})^2\right) {\rm d}x \ .
$$
Here $P$ denotes the degeneracy of $\mu_{K+1}$ and $G$ plays
the same role for $V-2F^{\prime}_c$ as $F$ does for $V$.
Although tempting, one cannot remove the cutoff in this 
formula since  the two terms on the right side are not separately continuous.
E.g., the degeneracy of the eigenvalue $\mu_{K+1}$ is not
necessarily the same as the degeneracy of the eigenvalue $\lambda_{K+1}$.
Nevertheless we have the following
$$
\sum_{j=1}^L \lambda_j^{3/2} - {3 \over 16}\int
{\rm Tr}(V^2) {\rm d}x =
$$
$$
\sum_{j=K+1}^L \lambda_j^{3/2} - {3 \over 16}\int
{\rm Tr}\left((V-2F^{\prime})^2\right) {\rm d}x =
$$
$$
\sum_{j=K+P+1}^L \mu_j^{3/2} - {3 \over 16}\int
{\rm Tr}\left((V-2F^{\prime}_c -2G^{\prime})^2\right) {\rm d}x \ + e_1 ,
$$
where $e_1$ is the error in the eigenvalues and the
potential integral due to the cutoff in the potential.
Again, the new potential has exponential decay.

By repeating the cutting and removing procedure
finitely many, say $s \leq L$  times, we end up with
$$
\sum_{j=1}^L \lambda_j^{3/2} - {3 \over 16}\int
{\rm Tr}(V^2) {\rm d}x =
 -{3 \over 16} \int {\rm Tr} (W^2) {\rm d}x + e_1 + \cdots +e_s \ ,
 \eqno(25) 
$$
where the $e_i$ denotes the error stemming from the cutoff
at the $i$-th step and $W$ is the resulting potential.
In particular (25) implies that
$$
\sum_{j=1}^L \lambda_j^{3/2} - {3 \over 16}\int
{\rm Tr}(V^2) {\rm d}x  \leq  e_1 + \cdots e_s
$$
which we can make as small as we please.\hfill\lanbox
\bigskip
{\bf Acknowledgment:} It is a great pleasure to thank Elliott Lieb for 
illuminating discussions and to Gian-Michele Graf for pointing out an
error in the original manuscript. M.L. would like to thank the Department of 
Physics at P. U. Cat\'olica de Chile for its hospitality.
\vskip .5 true in
{\bf References}
\item{[AL]} Aizenman and M., Lieb, E.H., On semi--classical bounds for
eigenvalues of Schr\"odinger operators, Phys. Lett. {\bf 66A}, (1978), 427--429.
\item{[BF]} Buslaev, V.S. and Faddeev, L.D., Formulas for traces for a singular 
Sturm--Liouville differential operator, Dokl. AN SSSR, {\bf 132 N 1}, (1960),
451--454.
\item{[C]} Crum, M.M., Associated Sturm--Liouville systems, Quart. J. Math. Oxford
Ser. 2, {\bf 6}, (1955), 121--127.
\item{[DG]} Darboux, D., Sur une proposition relative aux \'equations lin\'eaires,
C.R.Acad. Sci. Paris, {\bf 94}, (1882), 1456--1459.
\item{[DP]} Deift, P.A., Applications of a commutation formula, Duke Math. J. {\bf
45}, (1978), 267--310.
\item{[FZ]} Zakharov, V.E. and Faddeev, L.D., Korteweg--de Vries equation: A
completely integrable Hamiltonian system, Funct. Anal. Appl. {\bf 5}, 
(1978), 18--27. Translated from the Russian original  in Funkts. Anal. i Ego Pril.
{\bf 5}, (1971).
\item{[G]} Gesztesy, F., A Complete Spectral Characterization of
the Double Commutator \hfill\break
Method, J. Funct. Anal. {\bf 117}, (1993), 401--446.
\item{[HLT]} Hundertmark, D., Lieb, E.H. and Thomas, L.E., A Sharp Bound for an
Eigenvalue Moment of the One-Dimensional Schroedinger Operator,
Adv. Theor. Math. Phys. {\bf 2},(1998), 719- 731 .
\item{[HLW]} Hundertmark, D., Laptev, A. and Weidl, T., New bounds on the Lieb--
Thirring constants, to appear.
\item{[J]} Jacobi, C.G.J., Zur Theorie der Variationsrechnung und der
Differentialgleichungen, J. Reine Angew. Math. {\bf 17}, (1837), 68--82.
\item{[LT]} Lieb, E. and Thirring, W., Inequalities for the moments of the eigenvalues
of the \hfill\break
Schr\"odinger Hamiltonian and their relation to Sobolev
inequalities, Studies in Mathematical
Physics, Essays in honor of Valentine Bargmann, Lieb E., Simon B. , Wightman A, eds.
Princeton Series in Physics, Princeton University Press, Princeton 1976.
\item{[LW]} Laptev, A., Weidl, T., Sharp Lieb--Thirring inequalities in high
dimension, Preprint, 1999.
\item{[S]}  Schmincke, U.--W., On Schr\"odinger's factorization method for
Sturm-Liouville operators, Proc. Roy. Soc. Edinburgh Sect. A  {\bf 80}
(1978), 67-84.

\end